\documentclass[shortnote,twocolumn]{jpsj2}
\usepackage{graphicx}

\def\runtitle{a sample}
\def\runauthor{Yoichi \textsc{Asada}, Keith \textsc{Slevin} and Tomi \textsc{Ohtsuki}}

\title{%
The chiral symplectic universality class
}

\author{%
Yoichi \textsc{Asada}\thanks{asada@presto.phys.sci.osaka-u.ac.jp},
Keith \textsc{Slevin} and
Tomi \textsc{Ohtsuki}$^{1}$
}

\inst{%
Department of Physics, Graduate School of Science, Osaka University, 1-1 Machikaneyama, Toyonaka, Osaka 560-0043, Japan\\
$^1$Department of Physics, Sophia University, Kioi-cho 7-1, Chiyoda-ku, Tokyo 102-8554, Japan}

\recdate{\today}

\abst
{
}

\kword{%
Anderson localization, chiral symmetry,
spin-orbit coupling
}

\begin{document}
\sloppy
\maketitle

\section{Chiral symmetry}
About a decade ago Gade \cite{gade93} investigated the effect of
chiral symmetry in a nonlinear sigma model (Gade model).
She showed that this symmetry causes the localisation length
and the density of state to diverge at the band center in 2D.
This behaviour at the band center is different
from that found in the ``standard'' universality classes
of Anderson localisation.

If a Hamiltonian $H$ anti-commutes with
a unitary matrix $U$
then $H$ has chiral symmetry.
In this case energy eigenvalues
occur in pairs of opposite sign
with pairs of eigenstates $\psi_n$ and $U\psi_n$.
When there is no diagonal disorder
the Hamiltonian may have chiral symmetry.
This occurs if we can divide 
the system into two sublattices
such that nearest neighbour sites
belong to different sublattices.
If this is possible,
the transformation $U$ which changes the sign of the wave function
only on one sublattice
anti-commutes with the Hamiltonian.
If the total number of lattice sites is odd, chiral symmetry
guarantees the existence of
zero energy eigenstate(s), or zero mode(s).
Such states, if they exist,
have special properties e.g.
they are supported only on one sublattice.
The precise number of zero modes
depends on the difference
of the number of sites in each sublattice\cite{brouwer02}.

One example of a system with chiral symmetry
is the random magnetic flux model.
Anomalous behaviour is observed at the band center\cite{furusaki99}.
The results indicate that the entire band is localised
except for a critical state at the band center.

\section{The SU(2) model}
Here we report a numerical investigation
of localisation
in the SU(2) model \cite{asada1} without diagonal disorder.
The Hamiltonian of the SU(2) model has time reversal symmetry
but spin rotation symmetry is broken by a random spin-orbit coupling.
Thus we expect that states will belong either to the standard 
symplectic class or the chiral symplectic class.
This model is different from the random magnetic flux
model in that the zero mode is surrounded by
a metallic phase, not by an insulating phase.

The Hamiltonian of the SU(2) model describes non-interacting electrons
on a simple square lattice with nearest neighbour
SU(2) random hopping
\begin{eqnarray}
H=-\sum_{\langle i,j \rangle,\sigma,\sigma'}
U(i,j)_{ \sigma \sigma'}
c_{i\sigma}^{\dagger}c_{j\sigma'}
\label{eq:hamiltonian}
\end{eqnarray}
where $c^{\dagger}_{i\sigma}$($c_{i\sigma}$)
denotes the creation (annihilation)
operator of an electron at site $i$ with spin $\sigma$.
We distribute hopping matrices randomly and independently
with uniform probability on the group SU(2):
\begin{eqnarray}
U(i,j)=
\left(
\begin{array}{cc}
{\rm e}^{{\rm i}\alpha_{ij}} \cos \theta_{ij}
& {\rm e}^{{\rm i}\gamma_{ij}} \sin \theta_{ij} \\
-{\rm e}^{-{\rm i}\gamma_{ij}} \sin \theta_{ij}
& {\rm e}^{-{\rm i}\alpha_{ij}} \cos \theta_{ij}
\end{array}
\right)
\end{eqnarray}
where $\alpha$ and $\gamma$ are uniformly distributed
in the range $[0,2\pi)$,
and $\theta$ distributed in the range $[0,\pi/2]$
according to the probability density,
$P(\theta){\rm d}\theta=
\sin (2\theta) {\rm d}\theta$.

The SU(2) model has a mobility edge at $E_c\simeq 3.253$ 
\cite{asada1}.
In what follows we focus attention on the properties of
the model near the band center at $E=0$.
We calculate the localisation length on a quasi-1D strip
and attempt to extrapolate to the 2D limit.

\section{Behaviour of the localisation length}

We consider a quasi-1D strip whose width is $L$
and calculate the localization length $\lambda$
at arbitrary energy $E$
with the transfer matrix method \cite{mackinnon}.
In the transverse direction we impose either fixed boundary conditions (FBC)
or periodic boundary conditions (PBC).
The Hamiltonian has chiral symmetry except
when PBC are imposed on a system whose width is odd.
In this latter case chiral symmetry is broken.
We analyse the dependence of the re-normalized 
localization length $\Lambda = \lambda/L$
on $L$.

\begin{figure}
\begin{center}
\includegraphics[scale=0.3,angle=270]{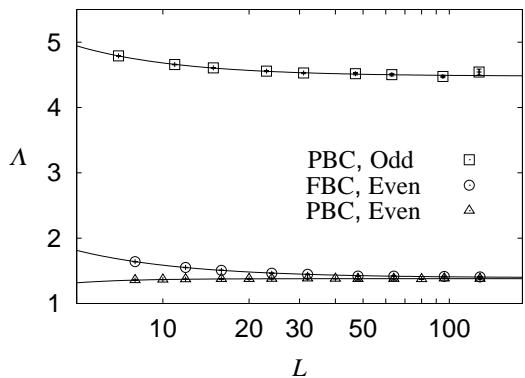}
\caption{$\Lambda$ vs $L$ at $E=0.0$.
The solid lines are the fit.}
\label{fig:loce0000}
\end{center}
\end{figure}

\begin{figure}
\begin{center}
\includegraphics[scale=0.3,angle=270]{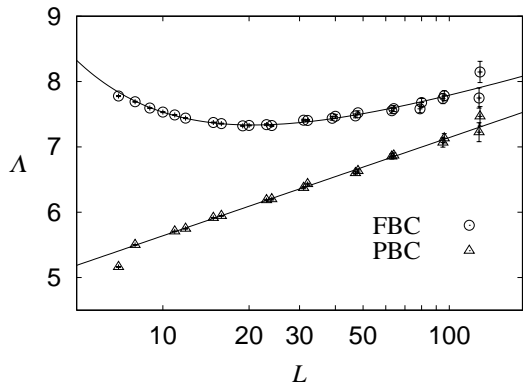}
\caption{$\Lambda$ vs $L$ at $E=0.5$.
The solid lines are the fit.
}
\label{fig:loce0500}
\end{center}
\end{figure}

The re-normalized localization length $\Lambda$
at $E=0.0$ and $E=0.5$
as a function of $L$
are shown in Figures ~\ref{fig:loce0000} and \ref{fig:loce0500}
respectively.
The width $L$ ranges from 7 to 128.
We performed $10^6 \sim 10^8$ transfer matrix multiplications
to obtain data
with an accuracy of $0.05\%$ to $2\%$.

Figure ~\ref{fig:loce0000} shows the results
at the band center $E=0.0$.
(Note that for odd $L$ and FBC the localisation length 
$\lambda$ diverges \cite{brouwer98} so no data is presented 
for this case.)
A striking dependence on the parity of $L$ is observed. 
As a function of $L$, $\Lambda$ approaches
a strongly parity dependent constant value for $L\rightarrow\infty$.
For even $L$ this asymptotic value is boundary condition  
independent.
To analyse the data in detail,
we fit the data to
\begin{equation}
\Lambda=a + c L^{-y}.
\label{eq:lambdae0}
\end{equation}
The results are tabulated in Table~\ref{table:fite0000}.
For even $L$ the estimates of the asymptotic value $a$
are the same within numerical accuracy.
For odd $L$ and PBC the asymptotic value is
much larger than for even $L$.
The $L$ independent behaviour of $\Lambda$ at large $L$
is typical of a critical point and may indicate that
band center is critical in this system.

In contrast at $E=0.5$ (Figure~\ref{fig:loce0500})
the parity dependence is negligible (except at
small $L$ under PBC).
Though a much strong boundary condition dependence is now
observed, it seems this will disappear in the limit of
large $L$.
For PBC the data can be fitted by
\begin{equation}
\Lambda=a+ b \ln L.
\label{eq:lambdapbc}
\end{equation}
and for FBC by
\begin{equation}
\Lambda=a+ b \ln L + cL^{-y}.
\label{eq:lambdafbc}
\end{equation}
The results are tabulated in Table~\ref{table:fite0500}.
In the limit of sufficiently large $L$ it seems that
a boundary condition independent logarithmic
increase of $\Lambda$ will be recovered.
Note in particular, by reference to Table~\ref{table:fite0500}, 
that the estimated (asymptotic) slopes for both
curves in Figure~\ref{fig:loce0500} are in agreement.

\begin{table}[tb]
\caption{
The best fit estimates of fitting parameters at $E=0.0$
with their 95\% confidence intervals.
$Q$ is the goodness of fit probability.
}
\label{table:fite0000}
\begin{tabular}{lllll} \hline
   & $Q$ & $a$ & $c$ & $y$ \\
\hline
FBC,Even&0.8&$1.39\pm .01$ & $2.6\pm .4$ & $1.12\pm .08$ \\
PBC,Even&0.8&$1.381\pm .002$ & $-2$ $(-6,-0.8)$ & $2.2\pm .5$ \\
PBC,Odd &0.8&$4.48\pm .03$ & $3.2\pm 1.3$ & $1.2\pm .2$ \\
\hline
\end{tabular}
\end{table}

To compare this behaviour with that in the metallic phase
of the standard symplectic class,
we also performed simulations of
the SU(2) model with diagonal disorder by
adding a term
\begin{equation}
H_d=\sum_{i,\sigma}\epsilon_i c_{i\sigma}^{\dagger}c_{i\sigma}
\end{equation}
to the Hamiltonian
(\ref{eq:hamiltonian}).
Here $\epsilon_i$ distributed randomly and independently
with uniform probability on the interval $[-1,1]$.
The diagonal disorder breaks chiral symmetry and
the system belongs to the standard symplectic class.
The re-normalized localization length $\Lambda$
is again a logarithmic function of $L$ 
and we are able to fit the result
by (\ref{eq:lambdapbc}).
We also get an estimate of the coefficient $b=0.64\pm .02$
in the SU(2) model which is consistent with
those in Table~\ref{table:fite0500} .
This indicates that
the states away from the band center
in the SU(2) model without diagonal disorder
are metallic and
belong to the standard symplectic universality class.

\begin{table}[tb]
\caption{
The best fit estimates of fitting parameters at $E=0.5$
with their 95\% confidence intervals.
$Q$ is the goodness of fit probability.
In fitting
we eliminated the data for $L\leq 8$.
}
\label{table:fite0500}
\begin{tabular}{llllll} \hline
  & $Q$ & $a$ & $b$ & $c$ & $y$ \\
\hline
PBC&0.1&$4.14\pm .04$ & $0.653\pm .013$ & & \\
FBC&0.2&$5.0\pm 1.6$ & $0.6\pm .3$ & $12\pm 6$ &$1.0\pm .3$ \\
\hline
\end{tabular}
\end{table}

\end{document}